\documentclass[final,3p,times]{elsarticle}

\usepackage{amssymb}

\journal{XXX}
\usepackage{setspace}
\usepackage[linesnumbered,ruled]{algorithm2e}
\usepackage{bm}
\usepackage{graphicx}
\usepackage{comment}
\usepackage{epstopdf}
\usepackage{caption}
\usepackage{subfigure}
\usepackage{float}
\usepackage{multirow}
\usepackage{booktabs}
\usepackage{booktabs}
\usepackage{array, threeparttable}
\usepackage{array}
\usepackage{amsmath}
\usepackage{booktabs}
\biboptions{numbers,sort&compress}
\usepackage[font=small,labelfont=bf,labelsep=none]{caption}
\usepackage[colorlinks=true,
linkcolor=black, 
anchorcolor=black, 
citecolor=black, 
urlcolor=blue
]{hyperref}
\begin{document}

\begin{frontmatter}



\title{Effective quality factor of mechanical resonators under complex-frequency excitations}

\author[lable1]{Wenbo Li}
\author[lable1,lable2]{Skriptyan Syuhri}
\author[lable1]{Pablo Tarazaga}
\author[lable1]{Raj Kumar Pal\corref{cor1}}
\ead{rkpal@tamu.edu}
\cortext[cor1]{Corresponding author}
\affiliation[lable1]{organization={J. Mike Walker '66 Department of Mechanical Engineering, Texas A\&M University},city={College Station},postcode={77843},state={Texas},country={USA}}
\affiliation[lable2]{organization={Department of Mechanical Engineering, University of Jember},city={Jember},postcode={68121},country={Indonesia}}

\begin{abstract}
We investigate the dynamics of mechanical resonators subject to excitations comprising of an oscillating or harmonic part, whose amplitude decays exponentially in time. We call these complex frequency excitations and show that the resulting response is quasi-steady, \textit{i.e.,} after an appropriate transform, the response of the new variable corresponds to the steady state behavior under a harmonic excitation. A procedure is presented to determine the amplitude-frequency response and effective quality factor based on this steady-state behavior. Optimal excitations are identified for both single and multi-degree of freedom systems that result in the amplitude-frequency response approaching that of an undamped system. The feasibility of the proposed method is verified through numerical simulations. 
Experiments with cantilever beams made of acrylic show a 54-fold increase in the effective quality factor. Our method does not involve any structural modifications and opens avenues for improving detection sensitivity in nondestructive testing and enhancing resolution in micro- and nano-electromechanical sensors.
\end{abstract}




\end{frontmatter}


\section{Introduction}
\label{Introduction}
The quality factor $Q$ of a  mechanical system at resonance plays a key role in multiple applications across length scales, from structural health monitoring (SHM) \cite{huang2020electromagnetically} of large-scale structures to micro- and nano-electromechanical (MEMS/NEMS)-based sensors~\cite{miller2018effective}. It is the ratio of the resonant frequency to its bandwidth under harmonic excitations and measures how sharply or rapidly the displacement varies with frequency near resonance.  For a fixed resonant frequency, $Q$ decreases with increasing damping ratio. In SHM applications based on resonance peak detection and frequency-shift identification, a high $Q$ increases resolution and improves the signal-to-noise ratio~\cite{rajasekaran2018free,lin2018applications}. On the other hand, a low $Q$ is often detrimental, since broader bandwidths make it hard to distinguish closely spaced frequency features, thereby complicating parameter identification. Furthermore,  damping typically causes adjacent modes to overlap in the spectrum  \cite{zhang2023vehicle}, making smaller modal frequency shifts difficult to detect \cite{limongelli2010frequency}. This issue is particularly pronounced in cracked structures, where a lower $Q$ renders methods based on output spectral analysis ineffective \cite{zeng2017dynamic,zhang2017vibration,bayma2018analysis}.
In MEMS/NEMS sensors, $Q$ is an important factor for evaluating the performance of a micro/nano-beam resonator \cite{hossein2006characterization,zhou2019single,dinh2024micromachined,qian2024broadband}. A low $Q$ substantially reduces sensing sensitivity and resolution, diminishes the signal-to-noise ratio, and increases susceptibility to noise \cite{rugar1991mechanical}. These factors underscore the importance of developing methods to improve the $Q$ factor for sensing applications.

There are two broad approaches to increasing the quality factor $Q$ of a resonant system. The first involves structural optimization or material modifications \cite{jain2014large,pollinger2009ultrahigh,ergincan2012influence,dania2024ultrahigh,mahashabde2020fast,chakram2014dissipation}. For instance, Geometry optimization strategies by Geerlings \textit{et al.} highlighted the potential of enhancing $Q$ through careful adjustments of resonator parameters \cite{geerlings2012improving}. Similarly, geometric alterations have been effectively utilized by Anatoliy \textit{et al.} to enhance resonant diaphragms' $Q$ \cite{kirilenko2014complicating}. Recently,  topology optimization has been successfully applied by Song \textit{et al.} to improve resonator performance \cite{song2022mems}. Additionally, surface modifications, like the gold coating applied by Sandberg \textit{et al.} on silicon dioxide microcantilevers, effectively modified their quality factors \cite{sandberg2005effect}. In electromagnetics, Sharma \textit{et al.} demonstrated that employing materials with varying dielectric constants significantly affects the resonators' quality factors \cite{sharma2008analysis}. 

The second approach involves managing external energy input or output pathways to increase a resonator's quality factor \cite{weld2006feedback,metzger2008self,okamoto2009controlling,taheri2017mutual,guo2013thermal}.  This approach does not alter the structure's geometry or intrinsic damping properties. The idea is to modulate the system's energy dissipation rate through feedback-based external energy input, for instance by applying an external force proportional to velocity. Under these conditions, the concept of an effective quality factor $Q_{eff}$ emerges~\cite{miller2018effective}. Examples include optical feedback methods by Kleckner \textit{et al.}, achieving enhanced micro-mechanical resonator performance by active optical feedback of atomic force microscope cantilevers \cite{kleckner2006sub}, and mechanical sideband excitation demonstrated by Venstra \textit{et al.}, enabling broad-range control of $Q_{eff}$ \cite{venstra2011q}. Other notable approaches at the small scales involve exploiting phonon-electron interactions, explored numerically, have significantly boosted resonator efficiency \cite{gokhale2014phonon} and using mechanical degenerate parametric amplification demonstrated experimentally to adjust micro-cantilevers' effective quality factors \cite{rugar1991mechanical}. Despite their theoretical promise, these externally driven methods frequently face practical implementation challenges or provide only limited enhancements to the effective quality factor, indicating a need for more robust and feasible strategies in resonator design.

Complex-frequency excitations, comprising of a harmonic part whose amplitude varies  exponentially, have recently emerged as a tool enabling novel and unique dynamic phenomena. The concept was first introduced by Archambault \textit{et al.}, who applied time-domain decaying excitations to a planar slab supporting surface plasmons, successfully overcoming resolution limitations caused by energy dissipation \cite{archambault2012superlens}. Since then, the concept has been extended to various fields, including photonics and elasticity,  leading to numerous breakthroughs. In photonics, Kim \textit{et al.} leveraged complex-frequency excitations to surpass conventional scattering response limitations, thus considerably enhancing the performance of optical devices \cite{kim2022beyond}. Guan \textit{et al.} advanced this approach further by utilizing harmonic measurements to synthesize complex-frequency excitation waves, achieving notable improvements in the imaging resolution of superlenses \cite{guan2023overcoming}. In elasticity, Rasmussen \textit{et al.},  demonstrated how complex-frequency shaping of input signals enables reflection-free and lossless absorption of incoming waves~\cite{rasmussen2023lossless}. Additionally, Loulas \textit{et al.} effectively employed complex-frequency incident radiation to enhance directional scattering characteristics in the terahertz regime \cite{loulas2024highly}. These studies collectively highlight the versatility and substantial potential of complex-frequency excitation methods in overcoming conventional resonator limitations. 

In this study, a method is proposed to increase the effective quality factor of an underdamped vibrating system by applying complex-frequency excitation. Our approach does not involve structural modifications nor does it require any feedback, in principle. This approach significantly improves the system’s effective quality factor, thus overcoming the aforementioned damping-induced limitations. The theoretical concept is presented using a linear spring mass damper system. Explicit expressions are derived for the required excitations that yield an effective quality factor $Q_{eff}$ approaching that of an undamped system. The concepts are extended to multi‑degree‑of‑freedom (MDOF), followed by an experimental demonstration with a continuous elastic structure, \textit{i.e.,}  a cantilever beam. 

The outline of the paper is as follows: first, we determine analytical expressions for complex-frequency excitations that yield a high $Q_{eff}$ and verify them through numerical simulations in Section \ref{Complex-frequency Excitation}. Then, we validate and demonstrate the effectiveness of the method through experiments in Sections \ref{Experimental Validation}. Finally, discussions and conclusions are presented in Section \ref{Conclusion}.

\section{Theory of Complex-frequency Excitations}
\label{Complex-frequency Excitation} 
We present a procedure that transforms a harmonic or narrow frequency band excitation into its complex frequency counterpart. Applying an appropriate change of variable, we show that the resulting behavior corresponds to the steady-state response under a harmonic excitation, which allows us to define an amplitude-frequency response and its effective quality factor. We derive conditions for the optimal complex frequency that maximizes this effective quality factor. Figure~\ref{fig2-1} displays a schematic of the concept. Applying a harmonic excitation $F_{ori}(t)$ results in a low quality factor for a damped system. In contrast, our approach achieves a higher effective quality factor by using a complex-frequency excitation $F(t)$. We then extend this concept to MDOF systems, followed by the verification of this concept with numerical simulations. 

\subsection{Complex-frequency and poles of transfer function}
\label{Complex-frequency Excitation for the single Degree-of-freedom System}
\begin{figure}[b!]
	\makeatletter
	\renewcommand{\fnum@figure}{Fig. \thefigure.\@gobble}
	\makeatother
	\centering
	\includegraphics[scale=0.42]{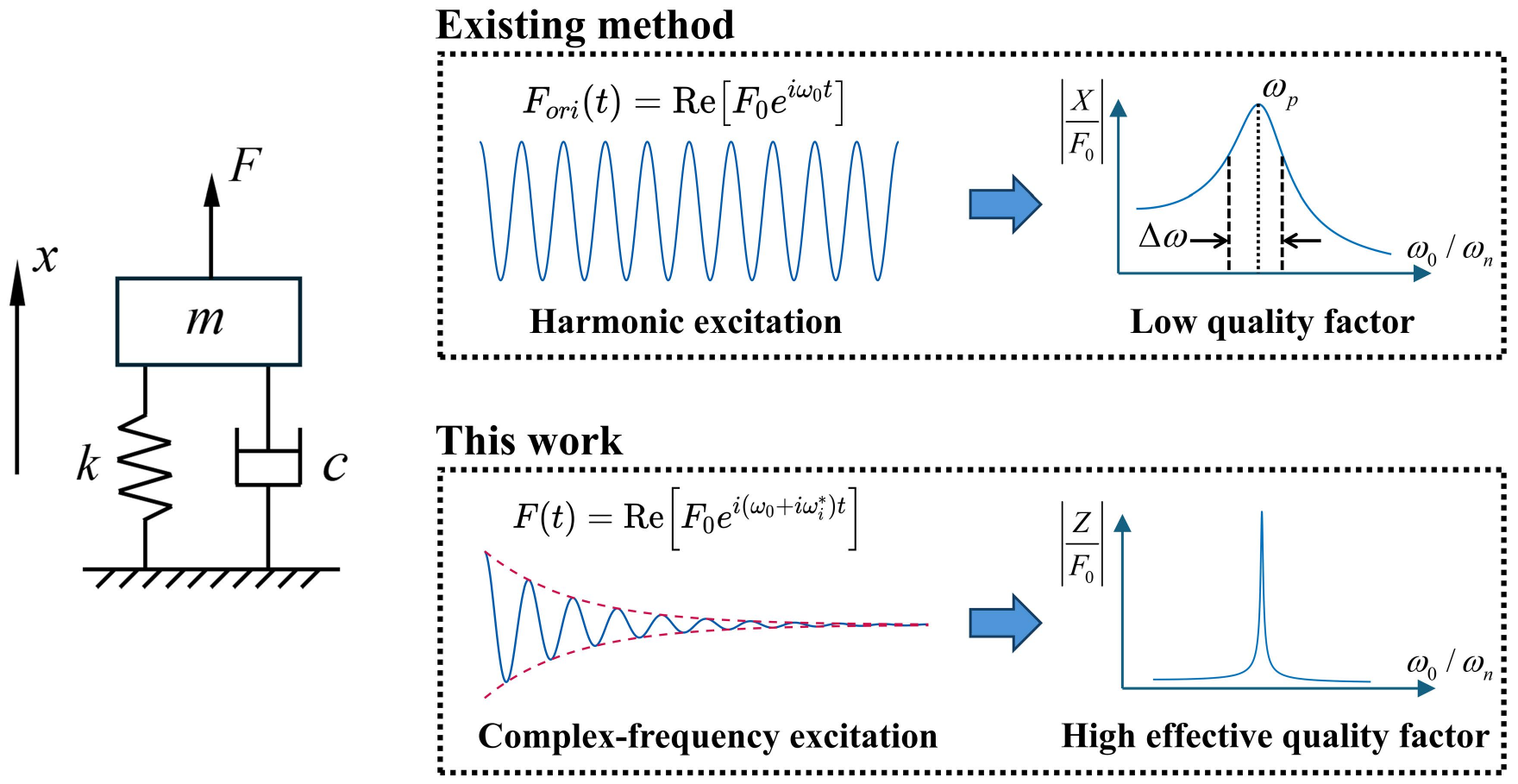}
	\caption{Overview of proposed concept. Under harmonic excitations, an under-damped system's frequency response curve is wide. Under complex‑frequency excitations, the curve becomes narrower, approaching that of an undamped system.} 
	\label{fig2-1}
\end{figure}

First, consider a SDOF linear spring–mass–damper system subjected to a harmonic force. The equation of motion can be written as
\begin{equation} \label{eq2-1}
	\ m\ddot{x}+c\dot{x}+kx={{F}_{0}}\sin (\omega t),
\end{equation}
where $x$ is the displacement of the mass along its degree of freedom, $m$ is the mass, $c$ is the damping coefficient, $k$ indicates the stiffness of the spring, $F_0$ is the amplitude of the excitation force, $\omega$ is the frequency of the excitation force, and $t$ is the time. For convenience, Eq.~\eqref{eq2-1} can be rewritten in complex form as 
\begin{equation} \label{eq:complex_sdof}
	\ m\ddot{x}+c\dot{x}+kx={{F}_{0}}{e^{i\omega t}},
\end{equation}
where $x(t)=X{e^{i\omega t}}$, and the imaginary part of $x(t)$ now gives the actual displacement. If $\omega$ is a real number, the steady-state response of this system at each frequency $\omega$ can be expressed in terms of a transfer function $H(\omega)$, given by
\begin{equation} \label{eq2-3}
	H(\omega) = \dfrac{X}{F_0}=\dfrac{1}{m(\omega _{n}^{2}-{{\omega }^{2}}+2i\zeta \omega {{\omega }_{n}})}
\end{equation}
and its magnitude is 
\begin{equation*}
    |H(\omega)| = \dfrac{1}{m\sqrt{(\omega_n^2 - \omega^2))^2 + 
    (2\zeta \omega \omega_n)^2}} . 
\end{equation*}
Here $\omega_n=\sqrt{k/m}$ is the system's natural frequency and $\zeta = c/(2m \omega_n)$ is its damping ratio. 
We work with underdamped systems, \textit{i.e.,} $0< \zeta < 1$. And the effective quality factor $Q_{eff}$ can be defined as
\begin{equation} \label{eq:Q_eff_defn}
	\ Q_{eff}=\frac{\omega_p}{\Delta \omega}=\frac{\omega_n\sqrt{1-2\zeta^2}}{\Delta \omega},
\end{equation}
where $\omega_p$ is the resonant frequency, and $\Delta \omega$ is the width of the frequency response at $1/\sqrt{2}$ of the peak amplitude, see inset in top box in Fig.~\ref{fig2-1}.

From Eq.~\eqref{eq2-3}, it can be observed that as the external force frequency $\omega$ approaches the natural frequency $\omega_n$, a smaller damping ratio $\zeta$ results in a sharper amplitude-frequency curve with a higher $Q_{eff}$. When the damping ratio $\zeta$ is zero, that is, when the system is undamped, $Q_{eff}$ becomes infinite. 
Note that $Q_{eff}\to 0$ when the denominator in Eq.~\eqref{eq2-3} goes to zero, \textit{i.e.,} at the poles of the transfer function $H(\omega)$.  At these points, $\omega$ satisfies
\begin{equation} \label{eq2-6}
	\ \omega _{n}^{2}-{{\omega }^{2}}+2i\zeta \omega {{\omega }_{n}}=0.
\end{equation}
If $\zeta \ne 0$, Eq.~\ref{eq2-6} does not admit any real solution. Our key idea is to allow $\omega$ to be a complex number, \textit{i.e.,}
\begin{equation} \label{eq2-7}
	\ \omega ={{\omega }_{r}^\ast}+i{{\omega }_{i}^\ast}.
\end{equation}
Then, by substituting Eq.~\ref{eq2-7} into Eq.~\eqref{eq2-6} and solving, we can obtain
\begin{equation} \label{eqn:complex_w}
	\ \left\{ \begin{array}{*{35}{l}}
   {{\omega }_{r}^\ast}=\pm {{\omega }_{n}}\sqrt{1-{{\zeta }^{2}}}  \\
   {{\omega }_{i}^\ast}=\zeta {{\omega }_{n}}  \\
\end{array} \right.
.
\end{equation}
We denote $\omega_r^*+ i{\omega }_{i}^\ast$ as the optimal complex frequency  and will demonstrate in Sec.~\ref{Quality Factor $Q$ under Complex Frequency Excitation} how it gives a narrower/sharper frequency response. When $\omega$ is complex, the force in Eq.~\eqref{eq:complex_sdof} becomes $F_0 e^{-\omega_i t} \sin(\omega_r t)$. This force excitation is exponentially decaying in time and $x(t)$ does not attain a steady state. Instead, as we discuss next, it attains a quasi-steady state, which allows the extraction of a frequency response and quality factor. 

\subsection{Quasi-steady state and effective quality factor}
Our objective is to get a sharper frequency response function centered around its resonant frequency, \textit{i.e.,} to get a higher $Q_{eff}$.  
To this end, we investigate the dynamic response of a system subject to a complex frequency excitation. Let us first illustrate how such excitations result in a quasi-steady state response. For a general complex frequency excitation with 
$\omega = \omega_r + i \omega_i $, Eq.~\eqref{eq2-1} may be written as 
\begin{equation}
    \ddot{x} + 2 \zeta \omega_n \dot{x} + \omega_n^2 x = 
    \dfrac{F_0}{m} e^{ {-\omega_i t}} \sin(\omega_r t) . 
\end{equation}
Performing a change of variable 
\begin{equation}\label{eqn:x-ztransf}
    x(t) = e^{{-\omega_i t}} z(t), 
\end{equation}
the above equation now becomes
\begin{equation}\label{eqn:z(t)}
    \ddot{z} + 2( \zeta \omega_n -\omega_i) \dot{z} + 
    (\omega_i^2 + \omega_n^2 - 2\zeta \omega_n \omega_i) z =\dfrac{F_0}{m} \sin(\omega_r t) . 
\end{equation}
Eq.~\eqref{eqn:z(t)} corresponds again to an SDOF system under a harmonic excitation. It has an effective damping constant $2( \zeta \omega_n -\omega_i)$, stiffness $(\omega_i^2 + \omega_n^2 - 2\zeta \omega_n \omega_i)$ and an effective unit mass. 
Thus under complex frequency excitations, the variable $z(t)$ attains a steady state response. The quasi-steady state of the actual displacement $x(t)$ is thus the steady state of $z(t)$ modulated by an exponential function $e^{-\omega_i t}$, see Eq.~\eqref{eqn:x-ztransf}. 
We call $\omega_i$ the decay exponent. Note that varying this decay exponent modifies both the effective damping ratio and natural frequency in Eq.~\eqref{eqn:z(t)}. In particular, we get a response similar to an undamped system when $\omega_i = \omega_i^* = \zeta \omega_n$. Thus, $\omega_i^*$ in Eq.~\eqref{eqn:complex_w} is hereby called the optimal decay exponent. Our idea is to characterize the frequency response of $z(t)$ as $\omega_r$ and $\omega_i$ vary, \textit{i.e.,} as the complex frequency varies.

\begin{figure}[!hbtp]
	\makeatletter
	\renewcommand{\fnum@figure}{Fig. \thefigure.\@gobble}
	\makeatother
    \makebox[\textwidth][l]{\hspace{1cm}\includegraphics[scale=0.43]{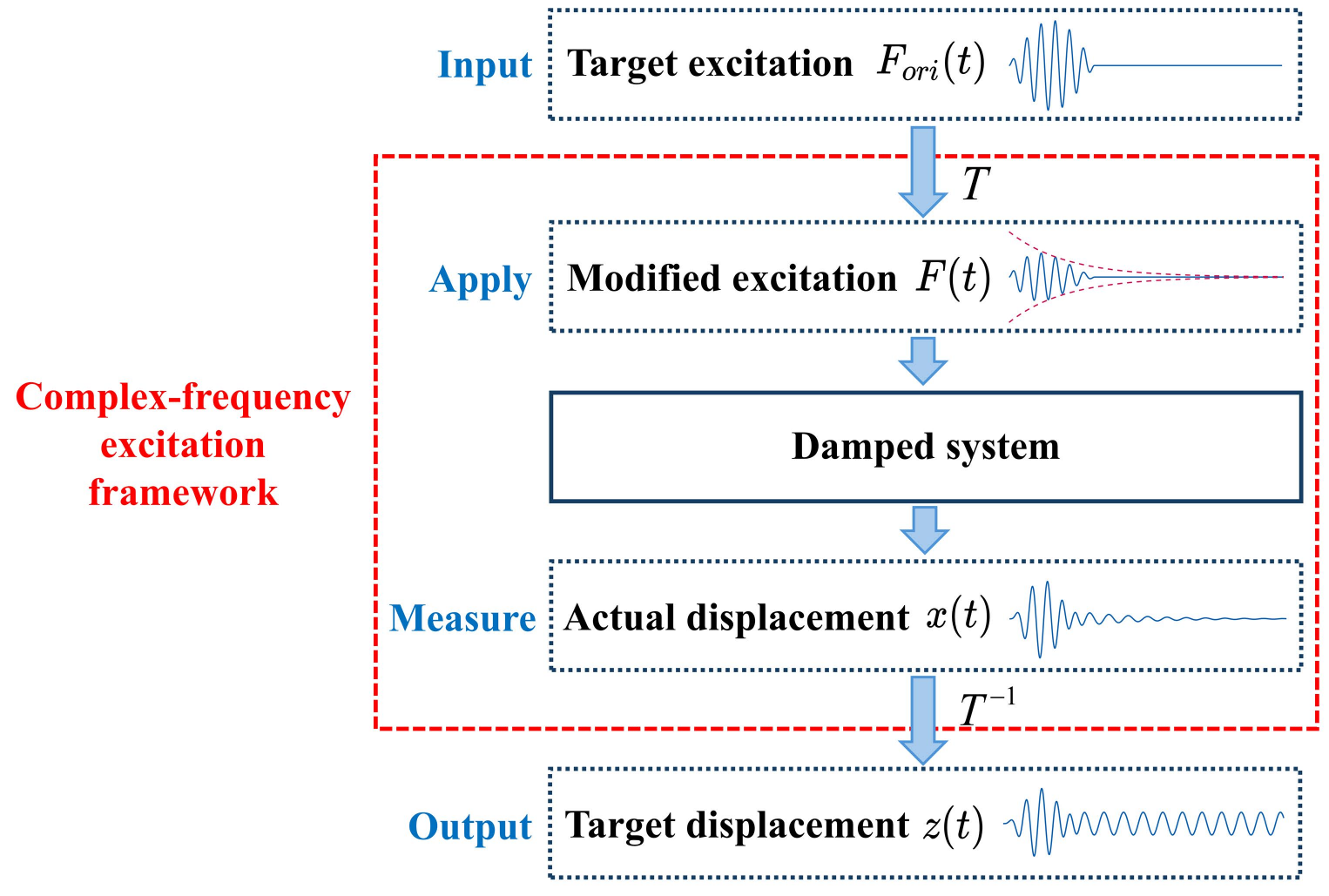}}
	\caption{Complex-frequency excitation procedure and framework. Operator $T$ modifies the target excitation $F_{ori}(t)$ and an excitation $F(t)$ is applied to the system. Its actual displacement is $x(t)$; applying operator $T^{-1}$ yields the target displacement $z(t)=T^{-1}[x(t)]$. Fourier transforms of $z(t)$ and $F_{ori}(t)$  are used to determine the frequency response and $Q_{eff}$.}	
	\label{fig2-2}
\end{figure}

The procedure to obtain this frequency response is outlined in Fig.~\ref{fig2-2}. Let $f(t,\omega)$ be a frequency dependent force excitation. This could either be a harmonic like $\sin(\omega t)$, or a windowed signal, like the one shown in Fig.~\ref{fig2-2}. Hereafter, we drop the index $\omega$ for brevity in $f(t,\omega)$ and call $f(t)$ the original or target excitation. Our goal is to track the response $z(t)$ as we vary the frequency $\omega$ in $f(t)$. We apply a modified excitation $F(t) $to the system, given by an operator $T_{\Omega}$ defined as 
\begin{equation}
  F(t) =   T_{\Omega}[f(t)] = e^{-\Omega t} f(t). 
\end{equation}
Here $\Omega$ is a real number. The inverse of this operator is defined as 
\begin{equation}
    T_{\Omega}^{-1}[f(t)] = T_{-\Omega} [f(t)] = e^{\Omega t} f(t) . 
\end{equation}

Note that this modified excitation $F(t)$ is the complex frequency ($\omega = \omega_r + i \omega_i$) counterpart of a harmonic excitation at frequency $\omega_r$. Indeed, for  $f(t) = F_0 \sin(\omega_r t)$, we have 
\begin{equation} \label{eq2-9}
	F(t)= \mathrm{Re}\left[ {{F}_{0}} {e^{i(\omega_r + i{{\omega }_{i}})t - i \pi/2}} \right] = F_0 e^{-\omega_i t} \sin (\omega_r t) = T_{\omega_i}\left[f(t) \right].
\end{equation}
We remark here that the resulting displacement $x(t)$ is broadband in frequency, in general, since $e^{\Omega t}$ is also broadband. However, the frequency spectrum of $x(t)$ is irrelevant for our purpose, since we instead seek the response of $z(t)$, given from Eq.~\eqref{eqn:x-ztransf} by 
\begin{equation}
    z(t) = T_{\Omega}^{-1} [x(t)] = e^{\Omega t} x(t) . 
\end{equation}
We call $z(t)$ the target displacement corresponding to the target excitation $f(t)$. The subscripts are hereafter dropped from $T$ for brevity. 
From Eq.~\eqref{eqn:z(t)}, we see that when $z(t)$ attains a steady state, its frequency spectrum is the same as that of $f(t)$. 
Based on the target excitation $f(t)$, target displacement $z(t)$ and their respective Fourier transforms, $F(\omega)$ and $Z(\omega)$, we now define a transfer function 
\begin{equation}
    H(\omega) = \dfrac{Z(\omega)}{F(\omega)}. 
\end{equation}
The effective quality factor $Q_{eff}$ in our proposed method is now computed using the peak and bandwidth of $H(\omega)$. 

To summarize, $Q_{eff}$ is determined from the target excitation and the target displacement. Here, $F(t)$ and $x(t)$ are the actual excitation applied to the system and its resulting displacement response, respectively, while $F_{ori}(t)$ and $z(t)$ may be regarded as the input and output to compute $Q_{eff}$. We will determine how the frequency response and $Q_{eff}$ changes with $\Omega$, \textit{i.e.,} the exponentially decaying part of a complex frequency excitation.

\subsection{Numerical Verification of effective quality factor}
\label{Quality Factor $Q$ under Complex Frequency Excitation}
To verify the effectiveness of the proposed method, we conduct numerical simulations on a SDOF damped spring–mass system. We apply a range of complex-frequency excitations and determine the transient response. The Newmark–$\beta$ method is used to solve for the actual displacement $x(t)$ of the mass, which then yields the target displacement $z(t)$. The initial conditions are chosen such that at $t = 0$, both the displacement and velocity of the mass are 0. Moreover, for each simulation, the total simulation time is set to 20 s, with a solution time step $\Delta t$ of $1\times10^{-4}$ s. Finally, the Fast Fourier Transform (FFT) is employed to extract the amplitude $Z$ of the target displacement $z(t)$ at each frequency, which then  yields the normalized amplitude–frequency response curve and the effective quality factor $Q_{eff}$ of the system under complex-frequency excitations. 

The equation of motion of the system under complex-frequency excitation can be written as
\begin{equation} \label{eq3-1}
	\ m\ddot{x}+c\dot{x}+kx={{F}_{0}}\cdot {e^{-{\omega }_{i}t}}\cdot \sin ({\omega }_{0}t).
\end{equation}
Here, \(\omega_i\) is the decay exponent and $\omega_0$ is the real or oscillating part of the complex-frequency. The parameter values chosen in Eq.~\eqref{eq3-1} are shown in Table \ref{tab1}.
\begin{table}[ht]
\renewcommand\arraystretch{1}
\centering
\caption{SDOF system parameters in the simulation.}
\label{tab1}
\begin{tabular}{llll}
\hline
Name                 & Symbol & Value & Unit   \\ \hline
Mass                 & $m$    & 1     & kg     \\
Stiffness           & $k$    & 10000 & N/m    \\
Damping coefficient & $c$    & 0.1   & $\text{N}\cdot \text{s}/\text{m}$ \\
Force amplitude      & $F_0$  & 1    & N      \\ \hline
\end{tabular}
\end{table}

For these values, the system’s natural frequency is $\omega_n= 100$ rad/s and its  damping ratio is $\zeta=0.05$; hence, $\omega_i^*=\omega_n \zeta= 5$ rad/s. To verify that $Q_{eff}$ is maximum when the decay exponent $\omega_i$ equals $\omega_i^{*}$, we conduct simulations for multiple distinct values of $\omega_i$. We vary $\omega_i$ from 0 rad/s to 10 rad/s in steps of 1 rad/s. Notably, when $\omega_i = 0$ rad/s, the excitation becomes the standard harmonic excitation. This implies that the amplitude–frequency response curve corresponding to $\omega_i = 0$ rad/s is precisely the system’s classical amplitude–frequency response curve under original excitation. To obtain a complete frequency response curve for each $\omega_i$, $\omega_0$ is varied independently from 10 rad/s to 200 rad/s in steps of 0.1 rad/s.

\begin{figure}[H]
	\makeatletter
	\renewcommand{\fnum@figure}{\textbf{Fig.} \thefigure.\@gobble}
	\makeatother
	\subfigure[]
	{
		\begin{minipage}[t]{0.5\linewidth}
			\centering
			\includegraphics[scale=0.53]{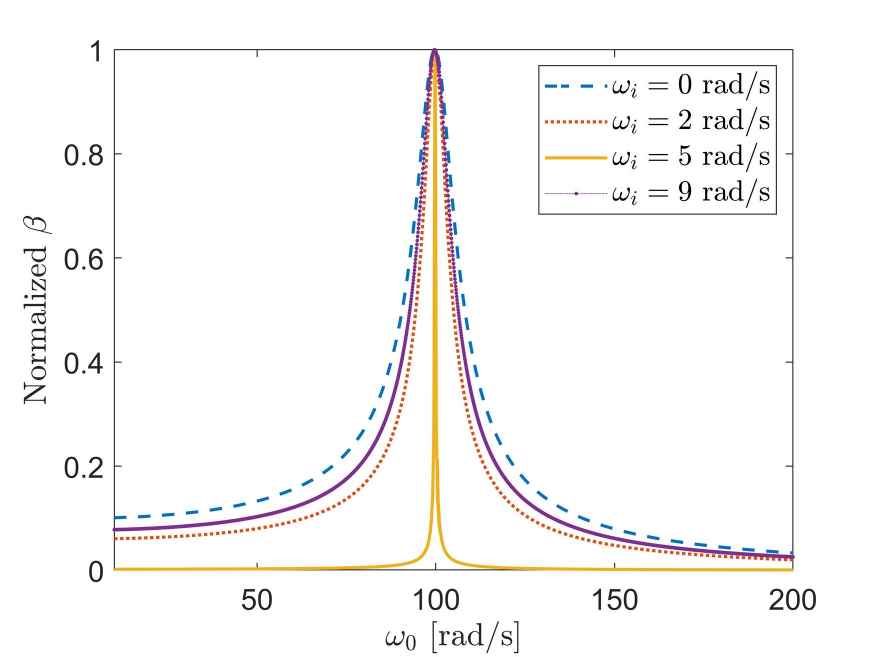}
			\label{fig3-1-a}
		\end{minipage}%
	}
	\subfigure[]
	{
		\begin{minipage}[t]{0.5\linewidth}
			\centering
			\includegraphics[scale=0.53]{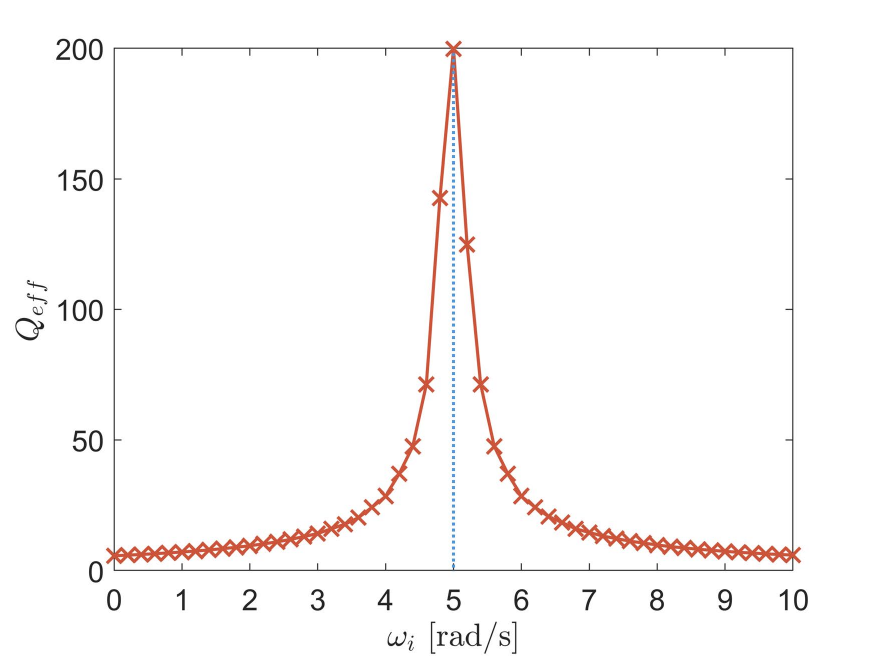}
			\label{fig3-1-b}
		\end{minipage}
	}
	\caption{ Numerical simulation results: (a) normalized amplitude–frequency response curves and (b) $Q_{eff}$ of the SDOF system. The response curves progressively sharpen as $\omega_i$ gets closer to $\omega_i^*$. $Q_{eff}$ attains its peak value precisely at $\omega_i=\omega_i^*$.}	
	\label{fig3-1}
\end{figure}

After obtaining the actual displacement $x(t)$ of the mass from transient simulations, the target displacement $z(t)$ can be obtained. Then, applying the FFT to $z(t)$ yields the amplitude $Z$ at that frequency $\omega_0$. The amplitude magnification factor $\beta$ is then obtained by
\begin{equation} \label{eq3-3}
	\beta=\frac{Z}{F_0}.
\end{equation}
Since the aim of this study is to determine the effective quality factor $Q_{eff}$ of the system's amplitude–frequency response, for ease of comparison, the complete amplitude–frequency response curve corresponding to each $\omega_i$ is normalized. The resulting normalized amplitude–frequency response curves are shown in Fig. \ref{fig3-1-a}.

From Fig. \ref{fig3-1-a}, we observe that when $\omega_i$ is at the optimal value of $\omega_i^*=5$ rad/s, the complex-frequency response curve becomes sharpest, which indicates the system attains its highest effective quality factor $Q_{eff}$ at this condition. To illustrate the relationship between $Q_{eff}$ and $\omega_i$ in a more intuitive and accurate manner while keeping all other parameters unchanged, we allow $\omega_i$ to increase from 0 rad/s to 10 rad/s in steps of 0.2 rad/s. For each $\omega_i$ value in this interval, we performed a transient simulation. The effective quality factor $Q_{eff}$ corresponding to each $\omega_i$ is shown in Fig. \ref{fig3-1-b}.

\begin{figure}[H]
	\makeatletter
	\renewcommand{\fnum@figure}{Fig. \thefigure.\@gobble}
	\makeatother
	\centering
	\includegraphics[scale=0.53]{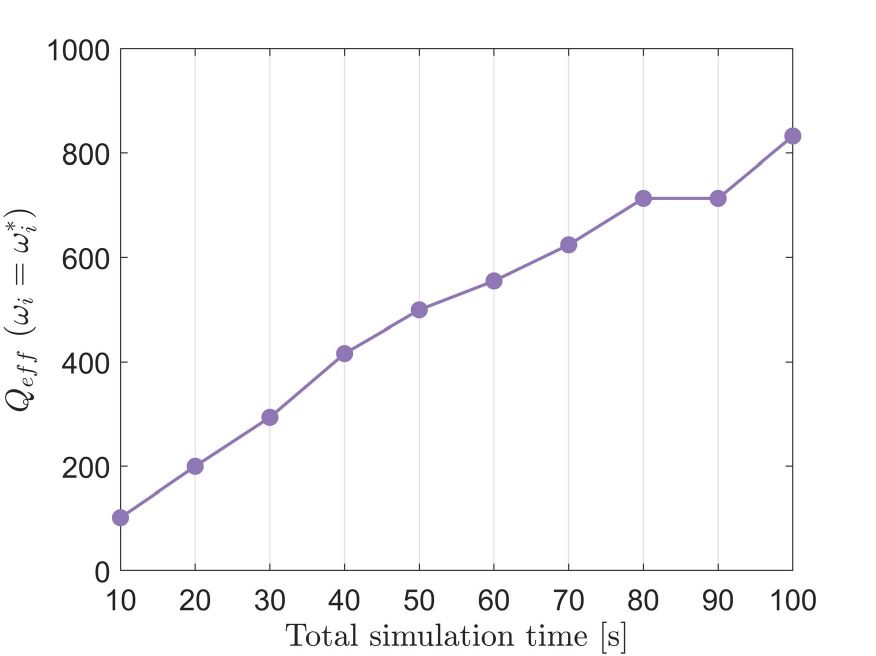}
	\caption{Computed $Q_{eff}$  at $\ (\omega_i = \omega_i^*)$ versus total simulation time $T_{final}$. Effective quality factor $Q_{eff}$ at the optimal decay exponent ($\omega_i=\omega_i^{*}$) increases with $T_{final}$, indicating that $Q_{eff}\to\infty$ as $T_{final} \to \infty$, similar to an undamped resonator.}	
	\label{fig2-7}
\end{figure}

Figure~\ref{fig3-1-b} shows that the effective quality factor $Q_{eff}$ reaches its maximum at $\omega_i= \omega_i^*=5\ \text{rad/s}$, which is in agreement with our theoretical predictions. Although the calculated \( Q_{eff}\ (\omega_i = \omega_i^*) \) at this finite simulation time remains finite, theoretically, its value approaches infinity. This discrepancy arises from the finite total simulation time used here; as the total simulation time approaches infinity, the maximum attainable $Q_{eff}$ tends to infinity. To verify this conclusion, we keep the other parameters constant and gradually increased the total simulation time from 10 s to 100 s in increments of 10 s. The resulting values of \( Q_{eff}\ (\omega_i = \omega_i^*) \) are shown in the Fig. \ref{fig2-7}. The results indicate that \( Q_{eff} \) increases continuously with the total simulation time, a characteristic of undamped systems.

\subsection{Actual Displacement and its maximum value}
\label{Maximum Actual Displacement under Complex-frequency Excitation}
To gain further insight on the nature of complex frequency excitations, let us examine a typical displacement vs time. We also determine the maximum displacement when $\omega_i = \omega_i^*$, \textit{i.e.,} at the pole of the transfer function. In a conventional undamped resonator with $\zeta = 0$, the response becomes unbounded at the resonant frequency and in reality, nonlinear effects dominate to alter the response. In contrast, we show here that the actual displacement is always bounded even at the poles when the transfer function $H(\omega)$ in Eq.~\eqref{eqn:x-ztransf} becomes unbounded. 

When the frequency $\omega$ of the external excitation equals the optimal one in Eq.~\eqref{eqn:complex_w}, the ratio of $Z$ to $F_0$ tends to infinity. However, we will show that the actual displacement $x(t)$ of the damped system remains bounded under an excitation of finite amplitude. When $\omega = \omega_r^{*} + i\omega_i^{*}$, the governing equation Eq.~\eqref{eqn:z(t)} for target displacement becomes 
\begin{equation*}
    \ddot{z} + (\omega_r^{*})^2 z = \dfrac{F_0}{m} \sin (\omega_r^{*} t).
\end{equation*}
Note that we have used the expressions for optimal complex frequency Eq.~\eqref{eqn:complex_w} to obtain the above equation. 
Its solution under zero initial conditions is 
\begin{equation}
    z(t) = \dfrac{F_0 t}{2 m \omega_r^{*}} \sin (\omega_r^{*} t) . 
\end{equation}
The actual displacement of the mass is then $x(t) = e^{-\omega^*_{i} t} z(t)$ and its amplitude $X(t)$ is 
\begin{equation} \label{eqn:X_ampl}
	X(t)= \dfrac{F_0 e^{-\omega_i t} }{2m\omega_{r}^*} t.
\end{equation}
Eq.~\eqref{eqn:X_ampl} shows that the amplitude $X(t)$ decays exponentially over time. Let us determine the maximum displacement amplitude and the corresponding time.
At that instant, the displacement satisfies
\begin{equation} \label{eq2-17}
	\dfrac{\mathrm{d}X(t)}{\mathrm{d}t}=0.
\end{equation}
This condiction gives the following maximum displacement amplitude $X_{\max}$ and its corresponding time $t_{\max}$:
\begin{equation} \label{eq2-18}
	X_{\max}=\dfrac{F_0}{2em\omega_r^* \omega_i^*},
\qquad
t_{\max}=\frac{1}{\omega_i^*}.
\end{equation}

Eq.~\eqref{eq2-18} indicates that at the optimal complex-frequency excitation, even though $|Z/F_0|$ goes to infinity, the actual amplitude $X$ does not increase indefinitely over time. Instead, there is an upper bound $X_{\max}$, attained around  $t = t_{\max}$. It is interesting to note that $t_{\max}$ is determined solely by $\omega_i^{*}$ and is independent of the target excitation frequency $\omega_0$.

To verify the correctness of the maximum actual displacement amplitude, we keep all parameters in the simulation same as that in Section \ref{Quality Factor $Q$ under Complex Frequency Excitation} and, following Eq.~\eqref{eqn:complex_w}, set $\omega_i=5 \ \text{rad/s}$ and $\omega_r=99.87 \ \text{rad/s}$. From Eq.~\eqref{eqn:X_ampl}, the theoretically predicted maximum actual displacement amplitude $X_{\max}$ is 0.367 mm at \(t_{\max}=0.2 \ \text{s}\). 
\begin{figure}[H]
	\makeatletter
	\renewcommand{\fnum@figure}{Fig. \thefigure.\@gobble}
	\makeatother
	\centering
	\includegraphics[scale=0.53]{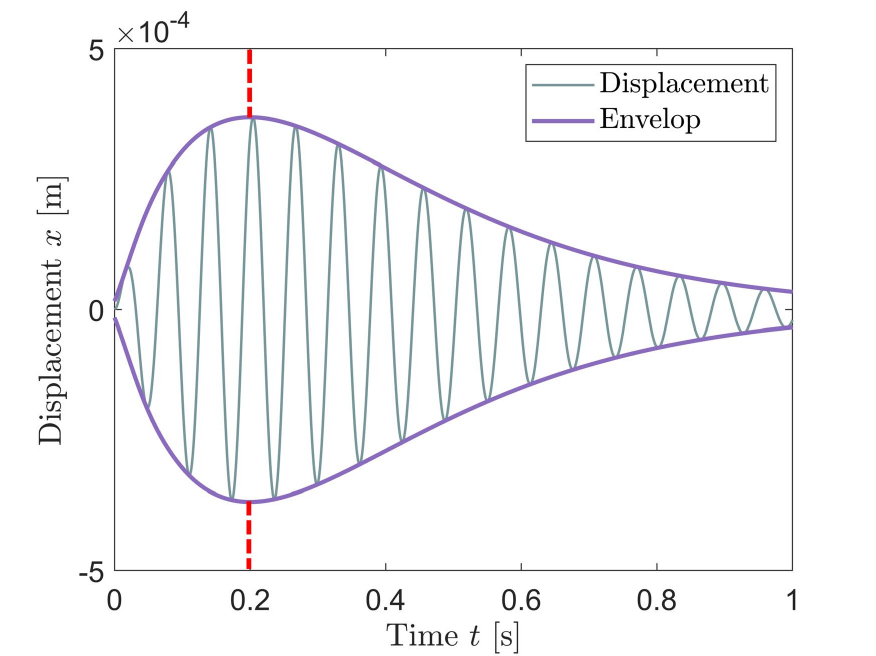}
	\caption{Actual displacement $x$ from transient simulation. The Hilbert envelope of $x(t)$ confirms the analytical bounds: maximum actual amplitude $X_{\max}\approx0.368$ mm occurs at $t_{\max}=0.2\;$s, consistent with $X_{\max}=F_{0}/(2e m\omega_r^{*}\omega_i^{*})$ and $t_{\max}=1/\omega_i^{*}$.}	
	\label{fig3-3}
\end{figure}

Fig. \ref{fig3-3} illustrates the actual displacement over the time interval \(0\!\sim\!1 \,\text{s}\) as obtained from the transient simulation. 
To determine the maximum displacement amplitude and its corresponding time, we performed a Hilbert transform \cite{marple1999computing} on the displacement signal to obtain its envelope, also shown in Fig. \ref{fig3-3}. From that envelope, we find the actual maximum displacement amplitude $X_{\max}^{(actual)}$ to be 0.368 mm at \(t_{\max}^{(actual)}=0.2 \ \text{s}\). These simulation results agree with our theoretical analysis and verify the accuracy of the derived expressions. 

\subsection{Complex-frequency Excitation for MDOF Systems}
\label{Complex Frequency Excitation for Multiple Degree-of-freedom System}
In Section \ref{Complex-frequency Excitation for the single Degree-of-freedom System}, we discussed how to determine $\omega_i$ for an SDOF system. Let us now discuss how to extend this concept to MDOF linear systems. The equations of motion of an $N$ degrees of freedom damped system may be expressed as
\begin{equation} \label{eqn:mdof}
\bm{M}\ddot{\bm{x}}+\bm{C}\dot{\bm{x}}+\bm{K}\bm{x}=\bm{F},
\end{equation}
where $\bm{M}$ is the $N\times N$ global mass matrix, $\bm{K}$ is the $N\times N$ global stiffness matrix, $\bm{C}$ is the $N\times N$ global damping matrix, $\bm{x}$ is the $N\times 1$ global displacement vector, and $\bm{F}$ is the $N\times 1$ global force vector. For harmonic excitations, $\bm{X}$ and $\bm{F}$ can be expressed in terms of their amplitudes $\bm{x_0}$ and $\bm{F_0}$ as
$	\bm{x}=e^{i\omega t}\bm{x_0}$ and $ \bm{F} = e^{i\omega t}\bm{F_0} $. 
The governing equation becomes 
\begin{equation} \label{eq2-22}
    (-\omega^{2}\bm{M}+i\omega\bm{C}+\bm{K}) \bm{x_0} = \bm{F_0}.
\end{equation}
The poles of this equation, \textit{i.e.,} the $\omega$ values at which $\bm{x_0}$ becomes unbounded for a finite $\bm{F_0}$ are given by setting $\det(-\omega^{2}\bm{M}+i\omega\bm{C}+\bm{K}) =0$, or equivalently, by solving the polynomial eigenvalue problem: 
\begin{equation} \label{eq2-24}
	(-\omega^{2}\bm{M}+i\omega\bm{C}+\bm{K}) \bm{x_0} = \bm{0}.
\end{equation}
Its solution gives $2N$ complex valued frequencies and $N$ mode shapes. These frequencies come in pairs of the form $\pm \omega_r^{(k)*} + i \omega_i^{(k)*}$, with $1 \leq k \leq N$, similar to the 1-DOF case in Eq.~\eqref{eqn:complex_w}. Similar to the single DOF system discussed earlier, the amplitude-frequency response is narrow and the effective quality factor is high when the system is subject to complex frequencies of the form $e^{-\omega^{(k)*}_i t} \sin(\omega t)$ and attains its peak at each of the $\omega^{(k)*}$. Note that the optimal complex frequencies are distinct from the natural frequencies of the system. The latter are obtained by solving the eigenvalue problem $(-\omega^2 \bm{M} + \bm{K}) \bm{x_0}= \bm{0}$.

\subsection{Numerical Verification for MDOF Systems}
\label{Simulation Verification for MDOF uner Complex Frequency Excitation}
\begin{figure}[H]
	\makeatletter
	\renewcommand{\fnum@figure}{Fig. \thefigure.\@gobble}
	\makeatother
	\centering
	\includegraphics[scale=0.3]{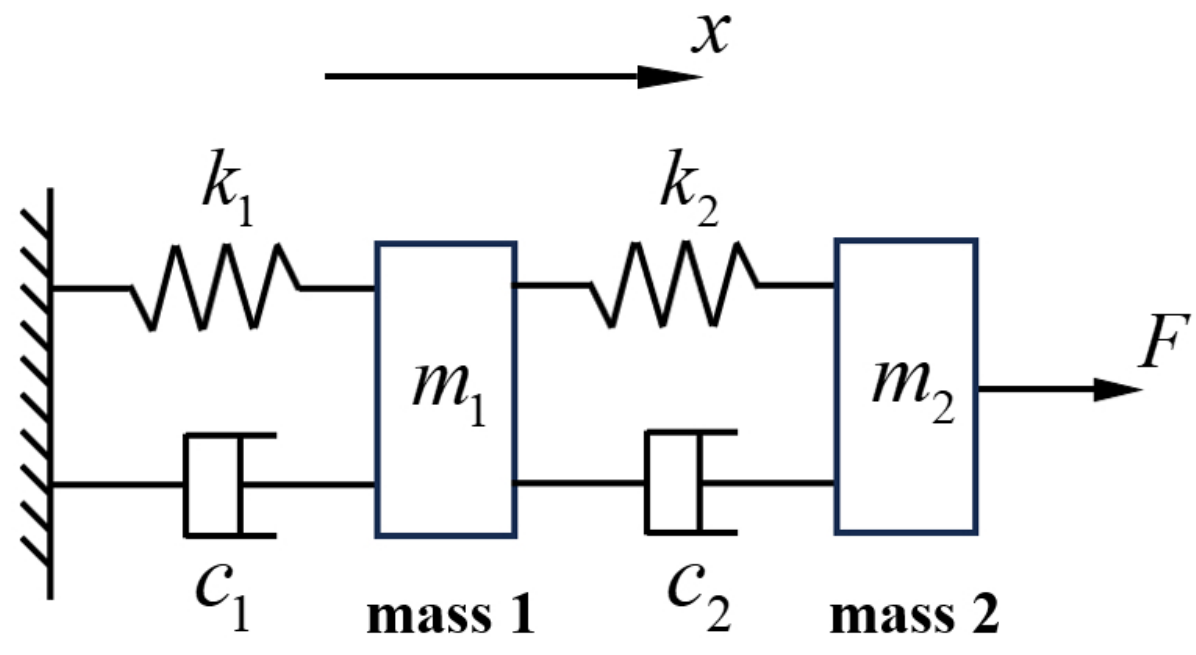}
	\caption{MDOF system in the simulation. Two‑degree‑of‑freedom spring–mass–damper model used for MDOF verification; a complex‑frequency force acts on $m_2$ to excite both modes.}	
	\label{fig2-3}
\end{figure}

Let us verify the concept using numerical simulations of the 2-DOF system shown in Fig.~\ref{fig2-3}. Complex-frequency excitations at various frequencies are applied to mass 2. Subsequently, the displacements of both masses are solved using the central difference method. The Newmark-$\beta$ method is not employed in this simulation because it introduces significant computational errors, rendering accurate displacement solutions unattainable. In this simulation, we focus primarily on the displacement of mass 2, extracting the amplitude of its target displacement through the FFT to obtain the normalized amplitude–frequency response curve and the effective quality factor $Q_{eff}$.

For the system in Fig.~\ref{fig2-3}, the various terms in the governing equation Eq.~\eqref{eqn:mdof} are 
\begin{equation} \label{eq2-28}
	\mathbf{M}=\begin{bmatrix}
     m_1 & 0\\
    0 & m_2
\end{bmatrix}, \quad
	\mathbf{C}=\begin{bmatrix}
     c_1+c_2 & -c_2\\
    -c_2 & c_2
\end{bmatrix},
\quad
\mathbf{K}=\begin{bmatrix}
     k_1+k_2 & -k_2\\
    -k_2 & k_2
\end{bmatrix},
\quad
	\mathbf{x}=\begin{bmatrix}
    x_1 \\
    x_2
\end{bmatrix},
\quad
\mathbf{F}=\begin{bmatrix}
    0 \\
    {{F}_{0}}  {e^{-{\omega }_{i}t}} \sin ({\omega }_{0}t)
\end{bmatrix}.
\end{equation}
The values of the MDOF system parameters in Eq.~\eqref{eq2-28} are shown in Table \ref{tab8}.
\begin{table}[ht]
\renewcommand\arraystretch{1}
\centering
\caption{MDOF system parameters in the simulation.}
\label{tab8}
\begin{tabular}{llll}
\hline
Name                 & Symbol & Value & Unit   \\ \hline
Mass 1                & $m_1$    & 2     & kg     \\
Mass 2                & $m_2$    & 0.5     & kg     \\
Stiffness 1          & $k_1$    & 600 & N/m    \\
Stiffness 2          & $k_2$    & 3000 & N/m    \\
Damping coefficient 1 & $c_1$    & 4   & $\text{N}\cdot \text{s}/\text{m}$ \\
Damping coefficient 2 & $c_2$    & 0.4   & $\text{N}\cdot \text{s}/\text{m}$ \\
Force amplitude      & $F_0$  & 1    & N      \\ \hline
\end{tabular}
\end{table}

Based on the values in Table \ref{tab8}, the optimal complex frequencies are \( \omega^{*(1)} = 15.4089+0.7869i\ \text{rad/s} \) and \( \omega_i^{*(2)} = 86.9528+0.7131i\,\text{rad/s} \). The corresponding natural frequencies of the two modes are $\omega_1 = 15.4283\ \text{rad/s}$ and $\omega_2 = 86.9596\ \text{rad/s}$. We then do transient simulations of transient response when various complex frequency excitations are applied, each with a total simulation time of 20 s. Two sets of amplitude-frequency response curves are presented, corresponding to the two modes.

\begin{figure}[H]
	\makeatletter
	\renewcommand{\fnum@figure}{\textbf{Fig.} \thefigure.\@gobble}
	\makeatother
	\subfigure[]
	{
		\begin{minipage}[t]{0.5\linewidth}
			\centering
			\includegraphics[scale=0.53]{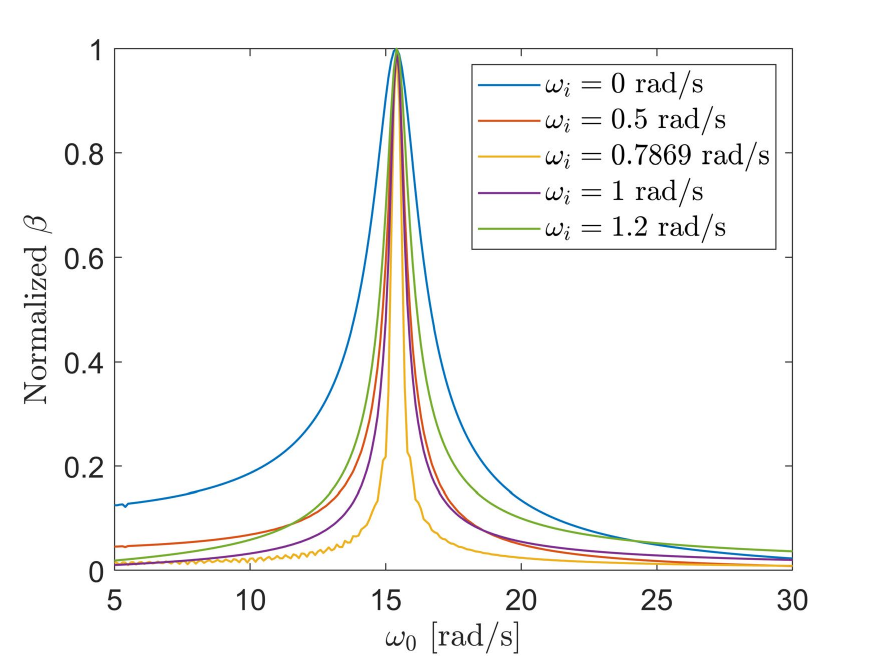}
			\label{fig2-4-a}
		\end{minipage}%
	}
	\subfigure[]
	{
		\begin{minipage}[t]{0.5\linewidth}
			\centering
			\includegraphics[scale=0.53]{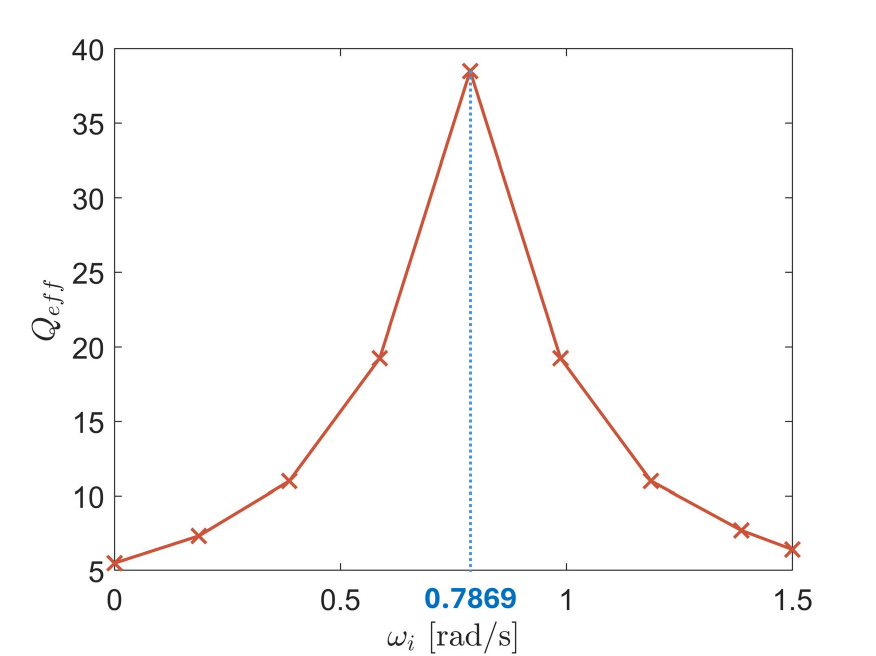}
			\label{fig2-4-b}
		\end{minipage}
	}
	\caption{(a) Normalized amplitude–frequency response curves and (b) $Q_{eff}$ of the first mode. Applying the optimal $\omega_i^{*(1)}=0.7869\ \mathrm{rad/s}$ yields the sharpest amplitude–frequency curve and the highest $Q_{eff}$.}
	\label{fig2-4}
\end{figure}

Figure~\ref{fig2-4-a} displays the amplitude-frequency response for five distinct values of the decay exponent $\omega_i$. The real or oscillating part of the complex frequency $\omega_0$ is varied from 5-30 rad/s in steps of \( 0.1\ \text{rad/s} \) to get the response around the first natural frequency. The curve is sharpest at \( \omega_i = \omega_i^{*(1)}= 0.7869\ \text{rad/s}\). Next, we determined the effective quality factor $Q_{eff}$ for this mode. Figure~\ref{fig2-4-b} displays how $Q_{eff}$ varies with $\omega_i$, clearly illustrating that the peak is attained at $\omega_i^{(1)*}$, the first optimal decay exponent.

\begin{figure}[H]
	\makeatletter
	\renewcommand{\fnum@figure}{\textbf{Fig.} \thefigure.\@gobble}
	\makeatother
	\subfigure[]
	{
		\begin{minipage}[t]{0.5\linewidth}
			\centering
			\includegraphics[scale=0.53]{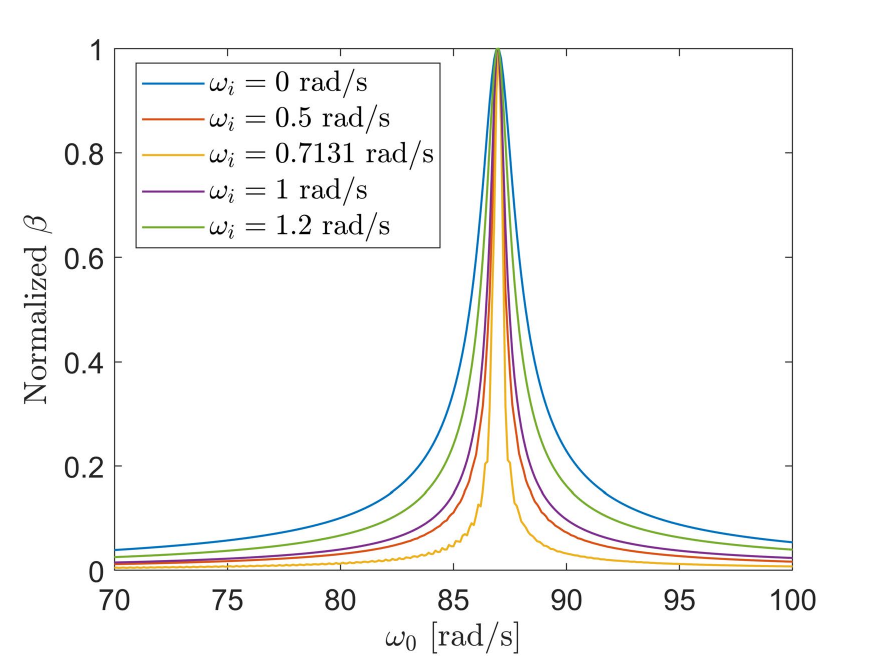}
			\label{fig2-5-a}
		\end{minipage}%
	}
	\subfigure[]
	{
		\begin{minipage}[t]{0.5\linewidth}
			\centering
			\includegraphics[scale=0.53]{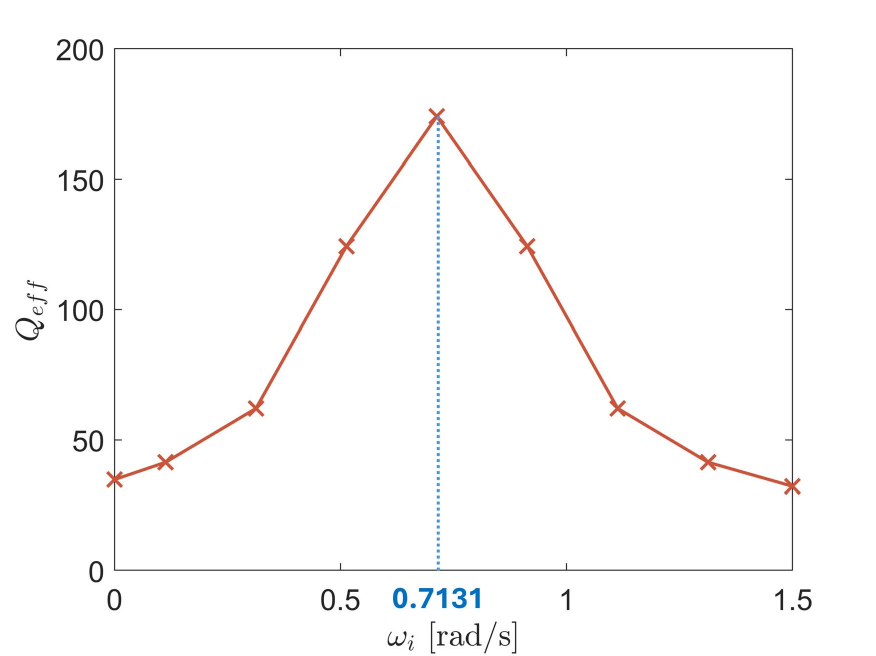}
			\label{fig2-5-b}
		\end{minipage}
	}
	\caption{(a) Normalized amplitude–frequency response curves and (b) $Q_{eff}$ of the second mode. Again, the decay exponent $\omega_i = \omega_i^{*(2)}=0.7131\ \mathrm{rad/s^{-1}}$  maximizes $Q_{eff}$, showing  mode‑specific nature of the optimal  complex‑frequency excitation.}
	\label{fig2-5}
\end{figure}

Similarly, for the second vibration mode, Fig.~\ref{fig2-5-a} displays the normalized amplitude-frequency response for five values of the decay exponent \( \omega_i \). Here, the oscillating part of the complex frequency varies from \( 70\ \text{rad/s} \) to \( 100\ \text{rad/s} \). Next, we analyzed the corresponding \( Q_{\text{eff}} \), and the results are shown in Fig. \ref{fig2-5-b}. Again, we make similar observations from Fig. \ref{fig2-5}. When the decay exponent equals the optimal value, \textit{i.e.,} \( \omega_i = \omega_i^{*(2)} \ \text{rad/s}\), the amplitude–frequency response curve of the second mode becomes the sharpest, and the effective quality factor \( Q_{eff} \) reaches its maximum value. These results verify the correctness and effectiveness of the complex-frequency excitation method for MDOF systems proposed in Section \ref{Complex Frequency Excitation for Multiple Degree-of-freedom System}.


\section{Experimental Validation with Cantilever Beam}
\label{Experimental Validation}
To demonstrate the feasibility of the proposed method, 
we implement complex‑frequency excitation on cantilever beams with contrasting damping levels—an aluminum beam (low damping) and an acrylic beam (high damping). A modal shaker delivers the programmed drive, while a force sensor and laser vibrometer provide synchronous force‑velocity measurements. Post‑processing with the inverse operator \(T^{-1}\) yields the target response required for \(Q_{eff}\) evaluation. Results show a 4.5‑fold increase in \(Q_{eff}\) for the aluminum beam and a 54‑fold increase for the acrylic beam when driven at their respective optimal \(\omega_i^{*}\), thereby validating the method across markedly different damping regimes and demonstrating its practical feasibility.

\subsection{Experimental Setup}
\label{Experimental Setup}
\begin{figure}[H]
	\makeatletter
	\renewcommand{\fnum@figure}{Fig. \thefigure.\@gobble}
	\makeatother
	\centering
	\includegraphics[scale=0.23]{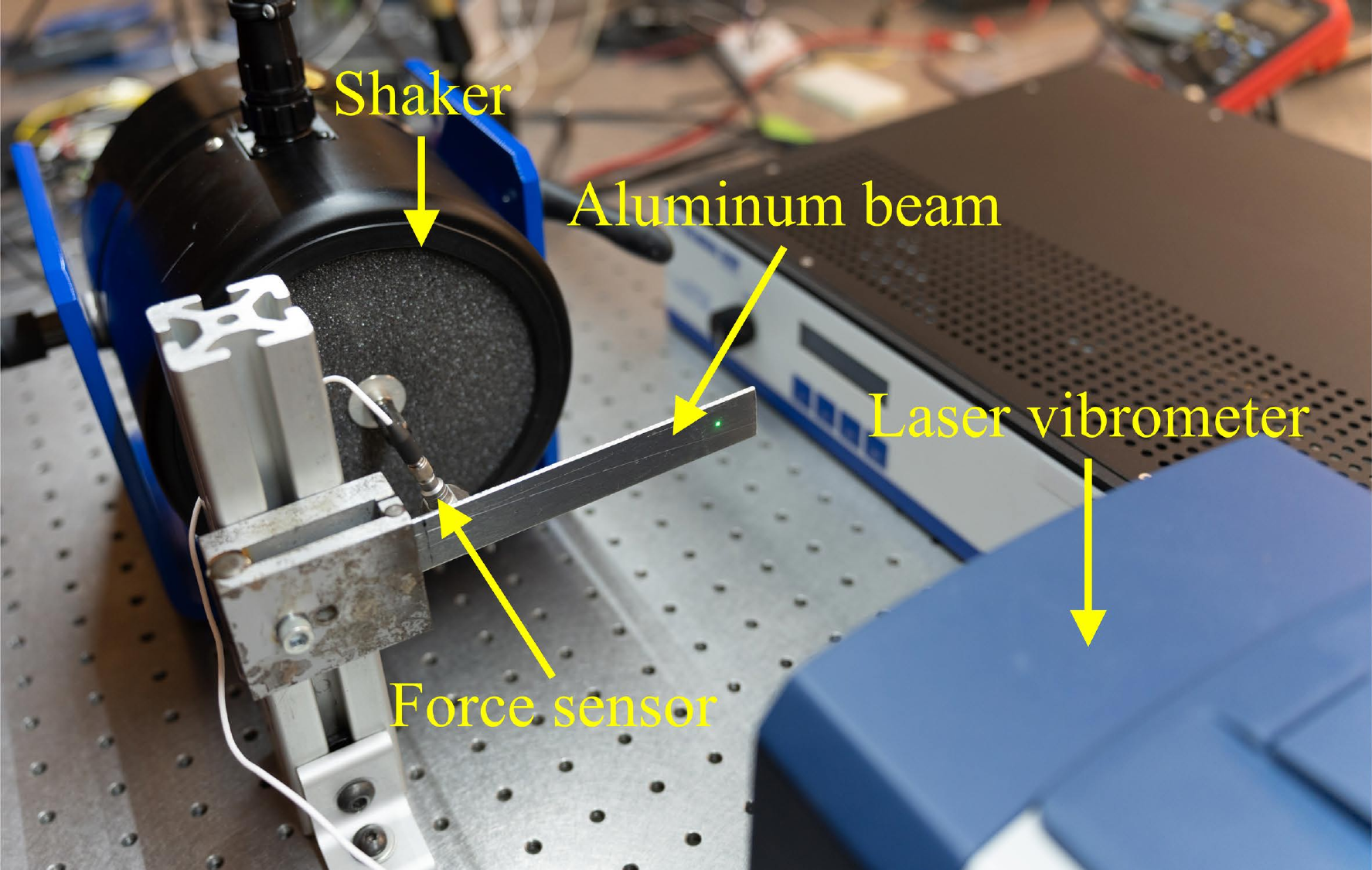}
	\caption{Experimental setup. The cantilever beam is excited by a shaker near its base. Its force at excitation is measured by a sensor attached to the shaker, while the tip displacement is measured by a laser vibrometer.}	
	\label{fig4-3}
\end{figure}

The experimental setup shown in Fig. \ref{fig4-3} is used to record and analyze the vibration of a cantilever beam under complex-frequency excitation. It comprises a fixture, a cantilever beam, an arbitrary waveform generator, a shaker, a force sensor, a laser vibrometer, and a data acquisition system. In this experiment, the excitation signal is generated via the Polytec software, which is integrated into the PSV-500 Polytec Scanning Laser Vibrometer (SLV) system. This system consists of the laser sensor, a dedicated signal processing unit with multiple input/output ports, and a software interface that controls both measurement and excitation functions. The arbitrary waveform generator outputs a complex-frequency signal which is then amplified using a SmartAmp 2100E21 Power Amplifier before being fed into the Modal Shop 2025E Modal Shaker to excite the cantilever beam.

We apply the following complex-frequency force excitation $F(t)$ to the cantilever beams:
\begin{equation} \label{eqn:expt_force}
	F(t) = e^{-\omega_i t} f(t) =\begin{cases}
   F_0 {e^{-{{\omega }_{i}}t}} \sin \left(\dfrac{\omega_{0}t}{N} \right)\sin (\omega_{0} t),
   & 0\le t\le {N\pi}/{\omega_{0}}  
   \\
   0, & {N\pi}/{\omega_{0}}<t\leq10. 
\end{cases}
\end{equation}
Here $F_0$ is constant and the excitation comprises of an $N$-cycle sine-windowed tone burst centered at frequency $\omega_0$ modulated by a decaying exponential $e^{-{{\omega }_{i}}t}$. 
Note that that we adopted the above excitation since our goal is to obtain the amplitude–frequency response curve of the beam in the vicinity of its natural frequency $\omega_n$. To reduce the number of experiments, we chose a signal with a certain bandwidth in the frequency spectrum. In this configuration, the chosen signal can be viewed as a superposition of multiple harmonic signals at different frequencies, which allows us to acquire the complete spectral response curve in a single experiment. In addition, because the excitation signal now spans a certain bandwidth in the frequency spectrum, there is no strict requirement for the value of $\omega_{0}$. It merely needs to remain sufficiently close to $\omega_{n}$.

As the cantilever beam vibrates under this excitation, the resulting velocity at the free end is measured using a PSV-500 Polytec SLV. The raw velocity data is transmitted to the data acquisition system, where they are preprocessed and stored. Simultaneously, the excitation force applied to the cantilever beam is measured using an ICP PCB Piezotronics 208C01 Force Sensor, which is mounted on the shaker and positioned close to the clamped section of the beam. These force measurements are likewise processed and stored for further analysis. We demonstrate the proposed concepts with beams made of two materials, which exhibit different damping properties, as test specimens.


\subsection{Case Study}
\label{Case Study}
\subsubsection{Aluminum Beam with Tape}
\label{Aluminum Alloy Beam}
Figure~\ref{fig4-3} shows the aluminum alloy Al 6061 beam. Its dimensions, material properties are given in Table \ref{tab2}. To increase its damping ratio, we applied tape to the beam’s surface, so that the effect of complex-frequency excitation on the beam’s effective quality factor can be more clearly demonstrated.
\begin{table}[ht]
\renewcommand\arraystretch{1}
\centering
\caption{Dimension and material properties of the aluminum alloy beam.}
\label{tab2}
\begin{tabular}{ccccccc}
\hline
Length (mm) & Width (mm) & Thickness (mm) &  Density (kg/$\text{m}^3$) & Young’s modulus (GPa) & Poisson's ratio \\ \hline
132.00      & 20.32      & 2.03           & 2700                      & 68                    & 0.33             \\ \hline
\end{tabular}
\end{table}

Before applying the complex-frequency excitation, we measure the natural frequency $\omega_n$ and damping ratio $\zeta$ of this cantilever beam in order to determine $\omega_i$. To this end, we conduct three repeated tests using the impact excitation method. The setup remains the same as in Fig. \ref{fig4-3}; only the modal shaker is replaced with a SAM NV-TECH-DESIGN Type 1 Modal Hammer. The hammer delivers controlled impact forces to the beam, while the resulting vibration response at the free end is captured using the Polytec SLV system. The Polytec software processes the force and velocity data to compute the frequency response function (FRF). The first natural frequency and damping ratio are then extracted from the FRF using curve fitting tools provided within the software.  The results from three repeated tests are presented in Table \ref{tab3}.
\begin{table}[ht]
\renewcommand\arraystretch{1}
\centering
\caption{Natural frequencies $\omega_n$ and damping ratios $\zeta$ of the aluminum alloy cantilever beam.}
\label{tab3}
\begin{tabular}{llllll}
\hline
           & Test 1 & Test 2 & Test 3 & Mean   & Standard deviation      \\ \hline
$\omega_n$ (rad/s) & 581.2   &  581.2   & 581.2    & 581.2    & 0                       \\
$\zeta$    & 0.0034 & 0.0038 & 0.0035 & 0.0036 & $2.0817 \times 10^{-4}$ \\ \hline
\end{tabular}
\end{table}

Based on Eq. \eqref{eqn:complex_w} and the results in Table \ref{tab3}, we get $\omega_i^*=\omega_n \zeta \approx 1.9\ \text{rad/s}$. 
We set $F_0 = 20\ \mathrm{N}$, $\omega_{al} = 565\ \mathrm{rad/s}$, and $N=48$ in Eq.~\eqref{eqn:expt_force}. Two sets of experiments are done: with $\omega_{i} = 0\ \mathrm{and}\ 1.9\ \mathrm{rad/s}$. We obtain the beam’s actual displacement $x(t)$ by numerically integrating the velocity data collected by the laser vibrometer. Using $x(t)$, we calculate the target displacement $z(t)$. Next, to obtain the amplitude–frequency response curve for computing the effective quality factor $Q_{eff}$, $z(t)$ must be transformed from the time domain into the frequency domain. The target displacement in the frequency domain, $Z(\omega)$, is obtained by discrete Fourier transform and is given by 
\begin{equation} \label{eq4-11}
	Z(\omega)=\sum_{n=0}^{N-1}z[n]\cdot  e^{-j\omega nT},
\end{equation}
where $N$ is the length of the target displacement data, and $T$ is the sampling interval.

Although the input force is prescribed, allowing a direct calculation of the amplitude corresponding to each excitation frequency, in practice various external factors due to noise result in the force applied to the beam not matching exactly with Eq.~\eqref{eqn:expt_force}. Therefore, we rely on the force sensor data $S(t)$ to determine  the target amplitude for each frequency. Similarly, we preprocess $S(t)$ to obtain the actual target excitation $f_a(t)$, which can be calculated as
\begin{equation} \label{eq4-2}
	f_a(t) = T(S(t)) = e^{\omega_i t} S(t).
\end{equation}
Next, we transform $f_a(t)$ in the time domain into the target excitation $F_a(\omega)$ in the frequency domain, again, using the discrete Fourier transform of $f_a(t)$. It is given by 
\begin{equation} \label{eq4-3}
	F_a(\omega)=\sum_{n=0}^{N-1}f_a[n]\cdot  e^{-j\omega nT}.
\end{equation}
Then, the amplitude magnification factor $\beta$ of the beam under complex frequency excitation is given by
\begin{equation} \label{eq4-4}
	\beta(\omega)=\left | \dfrac{ F_a(\omega)  }{ Z(\omega) } \right |.
\end{equation}

\begin{figure}[H]
	\makeatletter
	\renewcommand{\fnum@figure}{Fig. \thefigure.\@gobble}
	\makeatother
	\centering
	\includegraphics[scale=0.53]{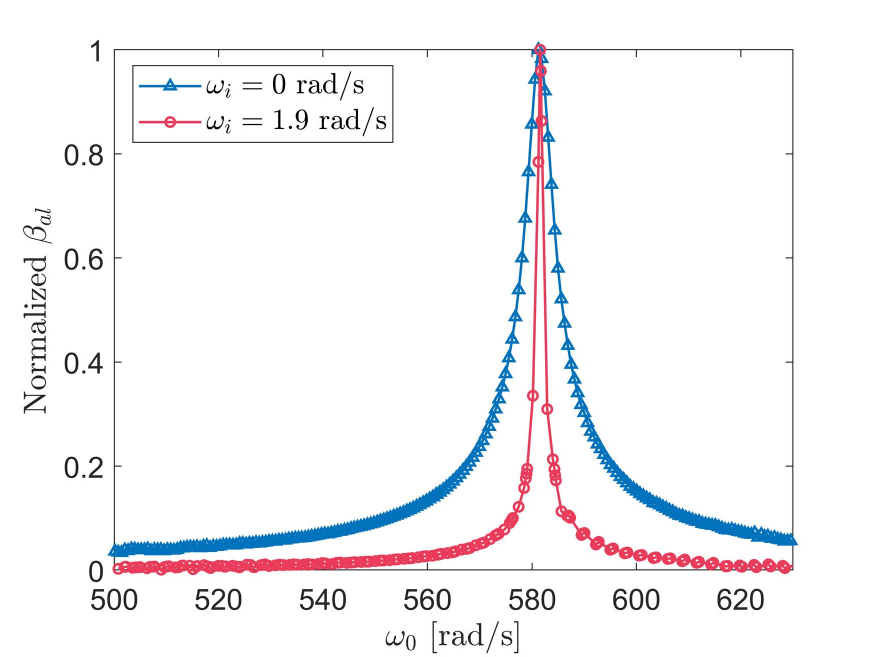}
	\caption{Amplitude–frequency response curves of the aluminum alloy cantilever beam. Complex frequency excitation with $\omega_i=1.9\ \mathrm{rad/s^{-1}}$ sharpens the amplitude–frequency response and boosts $Q_{eff}$ from $66.1$ to $294.8$.}	
	\label{fig4-1}
\end{figure}

Figure~\ref{fig4-1} displays the experimental amplitude–frequency response curves $\beta_{al}$ of the aluminum alloy cantilever beam under both standard and complex-frequency excitations. We observe that the response curve of the beam under complex-frequency excitation is more sharply peaked compared to that under a standard tone-burst excitation. Next, we determined the beam’s effective quality factor $Q_{eff}$ under each type of excitation. The values are presented in Table \ref{tab4}. These results indicate that the beam's $Q_{eff}$ significantly increases under complex-frequency excitation, becoming 4.5 times greater than under the original excitation, thus validating the effectiveness and feasibility of the proposed method.
\begin{table}[ht]
\renewcommand\arraystretch{1}
\centering
\caption{Effective quality factor $Q_{eff}$ of the aluminum alloy cantilever beam.}
\label{tab4}
\begin{tabular}{ccc}
\hline
    & Original excitation ($\omega_i = 0\ \text{rad/s}$) & Complex excitation ($\omega_i = 1.9\ \text{rad/s}$) \\ \hline
$Q_{eff}$ & 66.07                & 294.78                \\ \hline
\end{tabular}
\end{table}

\subsubsection{Acrylic Beam}
\label{Acrylic Beam}
\begin{figure}[H]
	\makeatletter
	\renewcommand{\fnum@figure}{Fig. \thefigure.\@gobble}
	\makeatother
	\centering
	\includegraphics[scale=0.23]{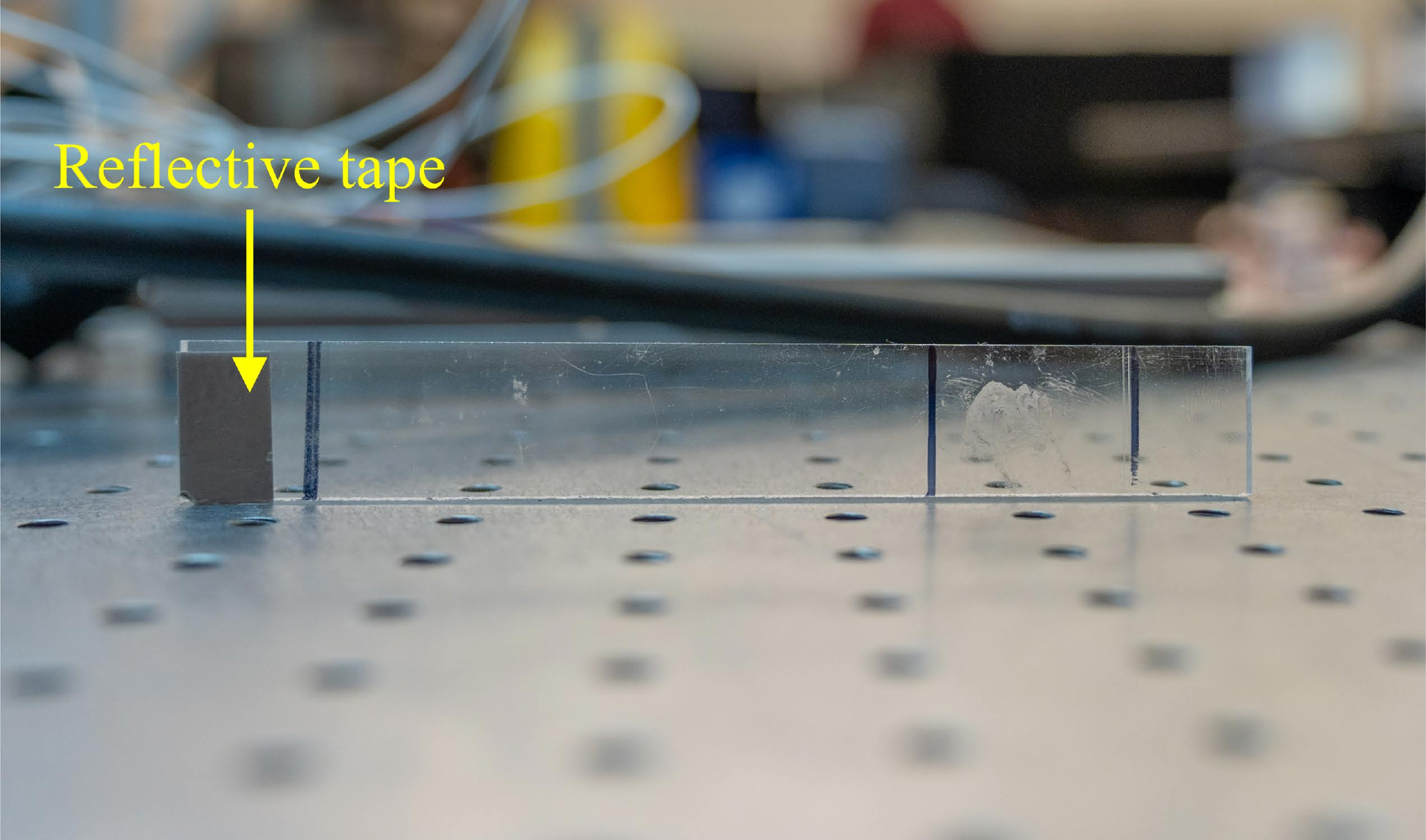}
	\caption{Acrylic beam. A reflective tape is attached so that the laser reflects off the otherwise transparent beam. }	
	\label{fig4-5}
\end{figure}

Next, we investigated the vibration response of a cantilever beam made of acrylic under both conventional and complex-frequency excitations. The purpose of this further analysis on a higher-damping system is to additionally validate the proposed method. See Fig. \ref{fig4-5} for an image of this beam. Its dimensions and material properties are listed in Table \ref{tab5}.

\begin{table}[ht]
\renewcommand\arraystretch{1}
\centering
\caption{Dimension and material properties of the acrylic beam.}
\label{tab5}
\begin{tabular}{ccccccc}
\hline
Length (mm) & Width (mm) & Thickness (mm) &  Density (kg/$\text{m}^3$) & Young’s modulus (GPa) & Poisson's ratio \\ \hline
130.00      & 20.00      & 4.76          & 1180                      & 2.8                    & 0.36             \\ \hline
\end{tabular}
\end{table}

Again, we employ the impact excitation method to measure the natural frequency $\omega_n$ and damping ratio $\zeta$ of the acrylic cantilever beam. The results are presented in Table \ref{tab6}. Consequently, we determine that $\omega_i^*=\omega_n \zeta \approx 15.6\ \text{rad/s}$ should yield the maximum possible effective quality factor. When the value of $\omega_i$ matches the theoretical value, the corresponding complex frequency excitation is termed the optimal  excitation; otherwise, it is referred to as a sub-optimal excitation.
\begin{table}[ht]
\renewcommand\arraystretch{1}
\centering
\caption{Natural frequencies $\omega_n$ and damping ratios $\zeta$ of the acrylic cantilever beam.}
\label{tab6}
\begin{tabular}{llllll}
\hline
           & Test 1 & Test 2 & Test 3 & Mean   & Standard deviation      \\ \hline
$\omega_n$ (rad/s) & 549.2   &  549.2   & 549.2    & 549.2    & 0                       \\
$\zeta$    & 0.02830 & 0.02807 & 0.02875& 0.02837 & $3.4588 \times 10^{-4}$ \\ \hline
\end{tabular}
\end{table}

Then, we apply the complex-frequency excitation in the form of Eq.~\eqref{eqn:expt_force}. The parameters are $F_0 = 20\ \mathrm{N}$, $\omega_{ac} = 534.1\ \mathrm{rad/s}$ and $N=24$. We do experiments for three distinct decay exponents, $\omega_{i} = 0,\ 7,\ \mathrm{and}\ 15.6\ \mathrm{rad/s}$. These values correspond to the standard tone burst ($\omega_i = 0$), an optimal ($\omega_i=15.7$ rad/s) and a sub-optimal ($\omega_i = 7$ rad/s) complex frequency excitation. 
The motivation is to investigate how $Q_{eff}$ changes when the excitation has an exponentially decaying component, but it is not the optimal one. 

\begin{figure}[H]
	\makeatletter
	\renewcommand{\fnum@figure}{Fig. \thefigure.\@gobble}
	\makeatother
	\centering
	\includegraphics[scale=0.53]{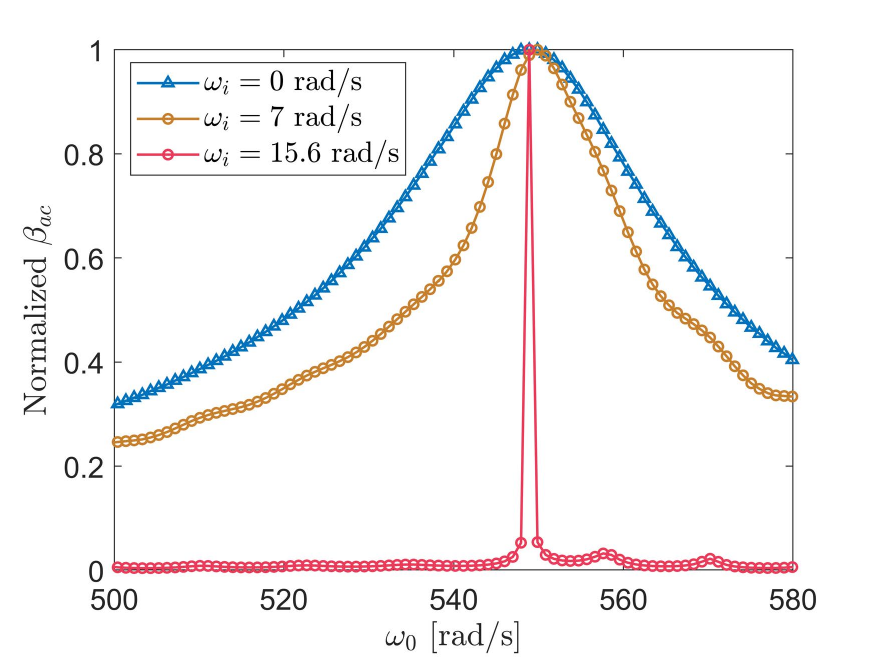}
	\caption{Amplitude–frequency response curves of the acrylic beam. Optimal complex‑frequency excitation ($\omega_i=\omega_i^*=15.6;\mathrm{rad,s^{-1}}$) raises the $Q_{\mathrm{eff}}$ 54 times, whereas a sub-optimal value ($\omega_i=7\ \mathrm{rad/s^{-1}}$) offers only a 1.7 times improvement, underscoring key role of decay exponent.}	
	\label{fig4-2}
\end{figure}

Using the acquired force data $F_{ac}$ and actual displacement data $x_{ac}$, we carry out processing and analysis in accordance with Eq.~\eqref{eq4-11}–\eqref{eq4-4}. This procedure yielded the amplitude magnification factor $\beta_{ac}$ for the acrylic cantilever beam. After normalizing $\beta_{ac}$, the results are shown in Fig. \ref{fig4-2}. Compared with the original excitation ($\omega_i=0$\ rad/s), complex-frequency excitations ($\omega_i=7,\ 15.6$\ rad/s) sharpen the amplitude-frequency response curves of the acrylic cantilever beam. Moreover, the optimal decay exponent ($\omega_i=15.6$\ rad/s) produces a significantly sharper amplitude-frequency response than the sub-optimal one ($\omega_i=7$\ rad/s). The frequency resolution $\Delta f$ of the Fourier transform here is $\Delta f = 1/T_{final} = 1/10 = 0.1$ Hz. Due to this resolution limitation, further refinement of the amplitude-frequency response curve under the optimal complex-frequency excitation is not feasible. Therefore, the effective quality factor $Q_{eff}$ is estimated based on the existing amplitude-frequency results, as shown in Table \ref{tab7}. It is anticipated that the actual value of $Q_{eff}$ under the optimal complex-frequency excitation will be greater than the estimated values derived from the current amplitude-frequency response curve. 

The results in Table \ref{tab7} indicate that, under optimal complex-frequency excitation, the effective quality factor $Q_{eff}$ of the acrylic cantilever beam is enhanced by 54 times compared to that under the original excitation. Under the sub-optimal complex-frequency excitation, the effective quality factor $Q_{eff}$ is 1.7 times that of the original excitation, although this improvement is significantly less than the optimal one. These findings demonstrate that the complex-frequency excitation method proposed in this study substantially enhances the effective quality factor of damped mechanical resonators, validating the effectiveness of the approach.

\begin{table}[htbp]
    \centering
    \caption{Effective quality factor $Q_{eff}$ of the acrylic cantilever beam for three excitations.}
    \label{tab7}
    {%
        \begin{tabular}{lccc}
        \toprule
         & Original excitation ($\omega_i = 0\ \text{rad/s}$)
         & Non-optimal  ($\omega_i = 7\ \text{rad/s}$)
         & Optimal ($\omega_i = 15.6\ \text{rad/s}$) \\
        \midrule
        $Q_{eff}$ & 10.48 & 18.29 & 566.16 \\
        \bottomrule
        \end{tabular}
    }
\end{table}

\section{Conclusion}
\label{Conclusion}
We investigated the dynamics of underdamped mechanical resonators subject to complex frequency excitations. Under a change of variable, the system exhibits a steady state response associated with the harmonic part of the complex frequency. The effective damping and thus the effective quality factor at this steady state depend on the exponential part of the complex frequency. As a result, we can modify the sharpness or width of the amplitude-frequency response by varying the complex frequency, i.e., the same system can exhibit distinct damping characteristics. 

For both single and MDOF systems, the complex valued poles of the transfer function are the optimal excitation frequencies at which the effective quality factor approaches that of an undamped system. In contrast to a conventional undamped resonator, whose amplitude becomes unbounded at its resonant frequency, the response under complex frequency excitations always remains bounded. Experiments with two cantilever beams of distinct materials showed significant improvement in their effective quality factors. Our results show the potential for enhancing the quality factor without requiring any structural modifications or feedback-based control. 

Finally, let us discuss some limitations and future research directions of our work. For underdamped systems with  high natural frequencies and large damping ratios, the corresponding decay exponent is high. This requires the excitation to undergo a very rapid attenuation over time, which is hard to generate in practice. In addition, extending this concept to over-damped and nonlinear systems will broaden its scope of application. We believe that addressing these challenges will lead to unraveling the full potential of complex frequency excitations for practical applications. 

\section*{Acknowledgment}
WL and RKP gratefully acknowledge support from NSF grant 2503599. 

\label{Acknowledge}






\bibliographystyle{elsarticle-num} 
\bibliography{bibref}

\end{document}